\documentclass[10pt,journal]{IEEEtran}
\usepackage[utf8]{inputenc}
\usepackage{amsmath}
\usepackage{amssymb}
\usepackage{graphicx} 
\usepackage{cuted, nccmath}
\usepackage{epsfig}
\usepackage{epstopdf}
\usepackage{mathbbol}
\usepackage{subfigure}
\usepackage{caption}
\usepackage{epsf}
\usepackage{bm}
\usepackage{amsthm, amsfonts, tikz, color, environ} 
\usepackage{stfloats}
\usepackage{float}
\usepackage{stmaryrd}
\usepackage[noadjust]{cite}
\usepackage{footnote} 
\usepackage{enumitem}
\usepackage{algorithm}
\usepackage{algorithmic}

\usepackage{array,multirow,makecell}
\setcellgapes{1pt}
\makegapedcells
\newcolumntype{R}[1]{>{\raggedleft\arraybackslash }b{#1}}
\newcolumntype{L}[1]{>{\raggedright\arraybackslash }b{#1}}
\newcolumntype{C}[1]{>{\centering\arraybackslash }b{#1}} 
\begin{document} 
\title{Integration of IRS in Indoor VLC Systems: Challenges, Potential and Promising Solutions}
\author{Mohamed Amine Arfaoui,
        Ali Ghrayeb, 
        and Chadi Assi
        \vspace{-0.7cm}
\thanks{M. A. Arfaoui and C. Assi are with Concordia Institute for Information Systems Engineering (CIISE), Concordia University, Montreal, Canada, e-mail:\{m\_arfaou@encs, assi@ciise\}.concordia.ca.}
\thanks{A. Ghrayeb is with the Electrical and Computer Engineering (ECE) department, Texas A$\&$M University at Qatar, Doha, Qatar, e-mail: ali.ghrayeb@qatar.tamu.edu.}}
\maketitle 
\thispagestyle{plain}
%-------------------------------------------------
\begin{abstract} 
Visible light communication (VLC) is an optical wireless communication technology that is considered a promising solution for high-speed indoor connectivity. Unlike the case in conventional radio-frequency wireless systems, the VLC channel is not isotropic, meaning that the device orientation affects the channel gain significantly. In addition, due to the use of optical frequency bands, the presence of different obstacles (e.g., walls, human bodies, furniture) may easily block the VLC links. One solution to overcome these issues is the integration of the intelligent reflective surface (IRS), which is a new and revolutionizing technology that has the potential to significantly improve the performance of wireless networks. IRS is capable of smartly reconfiguring the wireless propagation environment with the use of massive low-cost passive reflecting elements integrated on a planar surface. In this paper, a framework for integrating IRS in indoor VLC systems is presented. We give an overview of IRS, including its advantages, different types and main applications in VLC systems, where we demonstrate the potential of IRS in overcoming the effects of random device orientation and links blockages. We discuss key factors pertaining to the design and integration of IRS in VLC systems, namely, the deployment of IRSs, the channel state information acquisition, the optimization of IRS configuration and the real-time IRS control. We also lay out a number of promising research directions that center around the integration of IRS in indoor VLC systems.
\end{abstract} 
% - - - - - - - - - - - - - - - - - - - - - - - - - - - - 
\IEEEpeerreviewmaketitle 
\section{Introduction}
\subsection{VLC-Enabling 6G Networks}
\indent As the fifth generation (5G) cellular systems are currently under deployment, researchers from both academia and industry started shaping their vision on how the upcoming sixth generation (6G) would be \cite{saad2019vision}. 6G technologies are expected to go beyond patching the gaps and addressing the unfulfilled promises of 5G or keep up with the continuous rise of the Internet of-Things (IoTs) networks, 6G networks should also be able to handle the exponential increase of both the number of connected devices to the internet and the total data traffic. Therefore, these networks must urgently provide high data rates, seamless connectivity, ubiquitous coverage and ultra-low latency communications in order to reach the preset targets. \\ 
\indent In order to meet the high requirements of the upcoming 6G networks, researchers from both academia and industry are trying to explore new network architectures, new transmission techniques and new frequency spectra. One of the emerging technologies which has been proposed as a fundamental solution for 6G networks is visible light communication (VLC) \cite{strinati20196g}, which operates in the visible light frequency spectrum and uses light for both illumination and data communication purposes simultaneously. VLC has gained significant interest due to the high data rates this technology provides. The motivation behind the interest in VLC is twofold. First, the advantages that VLC offers when compared to RF, including the very large, unregulated bandwidth available in the visible light spectrum (more than 2600 times greater than the whole radio-frequency (RF) spectrum) \cite{VLC3}, the high signal transmission speed and enhanced security. Second, the availability of low cost off-the-shelf light-emitting-diode (LED) and photo-diode devices at the transmitter and receiver ends, respectively. 
\vspace{-0.3cm}
\subsection{Challenges and Existing Solutions}
\indent Unlike in conventional RF wireless systems, the VLC channel is not isotropic. This means that the orientation of VLC devices affects the communication performance significantly. Most of the studies on VLC
assume that the user equipment (UE) is perfectly aligned with the access point (AP), i.e., both the transmitter and the receiver are facing each other. However, such an assumption is not valid for the majority of smart communication devices. In fact, in real-life scenarios, the majority of users tend to hold their portable devices, such as wearable and smartphones, in a way that feels most comfortable. This means that the UE is not always facing the APs, and thus, can have an arbitrary orientation \cite{soltani2018modeling}. On the other hand, the blooming of IoTs has transformed many domains in our society, such as smart manufacturing. In this context, VLC technology has been shown to be a promising candidate for enabling industrial wireless networks, where seamless high speed connectivity can be provided between several machines, robots and sensors. These devices can be mobile and may also have random orientations during their movement and functioning when running under industrial processes. Such random orientation can significantly impact the operation of the industrial environment, which could affect the production quality and could even cause financial loss for factories. The second crucial factor, which can influence the performance of VLC systems is the blockage of wireless links. In fact, due to the use of the optical frequency bands, the VLC links may be easily blocked, either by the user themselves, known as self-blockage, or by other users or objects (e.g., furniture, walls, machines, etc.), and this can consequently interrupt the communication links and put the VLC systems in outage \cite{blockageIET}. \\ 
\indent Both device orientation and link blockage can significantly affect the performance of VLC systems, such as users' throughput and reliability. Only a few studies considered the impact of random orientation and blockage on the performance of indoor VLC systems and proposed solutions to alleviate their effects \cite{arfaoui2020measurements,mohammad2018optical}. In \cite{arfaoui2020measurements}, statistical VLC channel models were derived while considering the orientation randomness of VLC UEs. Then, an optimized design of VLC cellular systems was proposed to reduce the outage of VLC systems. On the other hand, a multi-directional configuration for VLC receivers was proposed in \cite{mohammad2018optical}, which was shown to be robust against the random orientation of VLC users and the blockage of optical links. Although some solutions were proposed in the literature \cite{arfaoui2020measurements,mohammad2018optical}, the aforementioned studies focused only on optimizing the transmitter and receiver architectures, and consequently, there is no control on the wireless propagation environment. This observation gives rise to the following question: "\textit{Is there a way to control and/or reconfigure the wireless propagation environment that can beat the effects of the random receiver orientation and link blockage in VLC systems?} The answer is in fact yes, and this is accomplished by using the intelligent reflective surface (IRS) technology.
\vspace{-0.3cm}
\subsection{The Need For IRS technology}
\indent IRS has been proposed as a promising new technology for reconfiguring the wireless propagation environment via software-controlled reflection \cite{wu2019towards}. Specifically, IRS is a planar surface comprising a large number of low-cost passive reflecting elements, each being able to reflect the signal into a specific direction. Hence, the reflecting elements are collaboratively achieving three-dimensional (3D) reflect beamforming, also known as passive beamforming, since no additional transmit power is required. Recently, IRS has received significant research attention for a number of areas, including wireless fidelity (Wi-Fi), millimeter wave (mmWave), terahertz (THz), and free space optical (FSO) communications, covering a large frequency range. The common observation is that IRS can be utilized to create desirable radiation patterns in challenging wireless propagation environments and, consequently, a large body of research and experimental works has been dedicated to model and study the potential gains of using such technology for these wireless communication systems \cite{gong2020toward}.\\
\indent As for VLC systems, IRS is expected to contribute effectively in boosting their performance especially that most of VLC systems rely on the presence of direct line-of-sight (LOS) links. In fact, in sharp contrast to existing wireless link adaptation techniques at the transmitter and the receiver proposed in \cite{arfaoui2020measurements,mohammad2018optical} and references therein, IRS can proactively modify the wireless channel between the VLC transmitters and receivers via highly controllable and intelligent signal reflections. Thus, IRS provides a new degree of freedom to further enhance the performance of VLC systems and paves the way to realize a smart and programmable optical wireless environment. Consequently, the incorporation of IRSs in indoor VLC systems can provide significant gains, especially in terms of resilience to the random orientation of the communication devices and to the blockages of communication links between the VLC transmitters and receivers.
\vspace{-0.3cm}
\subsection{Objectives and Outline}
IRS-aided VLC systems consist of both active components, such as APs and UEs, and passive components, which are the IRSs, thus they differ significantly from the traditional VLC systems comprising only active components. This is indeed the motivation of this article to provide an overview of IRS-aided VLC systems, including the deployment and integration of IRS, acquisition of the channel state information (CSI), passive beamforming design, real-time IRS control, to name a few. In particular, the main challenges and their potential solutions for designing and implementing IRS-aided VLC systems are highlighted to inspire future research directions. Finally, a number of promising research directions that center around the integration of IRS in indoor VLC systems are proposed.
\section{Types of IRS and Their Potential}
\subsection{Types}
\indent A particularly interesting application of IRS in VLC systems is the realization of optical power focusing toward target receivers \cite{bjornson2020reconfigurable}. Recently, efforts were made to investigate the performance of IRS for optical communication systems \cite{najafi2019intelligent,abdelhady2020visible}. In \cite{najafi2019intelligent}, smart mirrors were proposed to relax the LOS requirement for FSO links. Moreover, in \cite{abdelhady2020visible}, two types of IRS were proposed for VLC systems to focus the incident optical power toward a VLC receiver, where the first type is based on programmable metasurfaces and the second type is based on traditional mirrors. These two types of IRS are defined in the following. 
\begin{figure}[t]
\centering     
\subfigure[MSA element (metasurface patch).]{\label{fig:MSA_reflection}\includegraphics[width=0.51\linewidth]{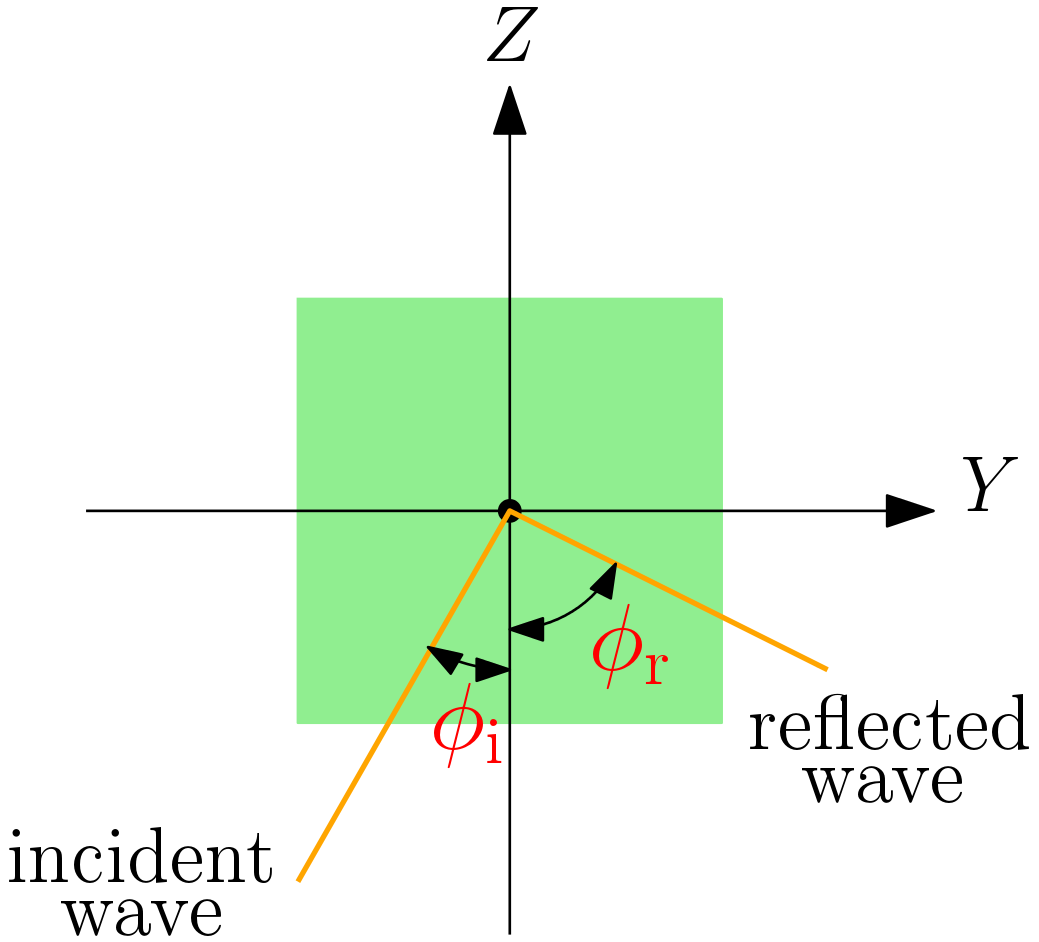}}%
\subfigure[MA element (small mirror).]{\label{fig:MA_reflection}\includegraphics[width=0.51\linewidth]{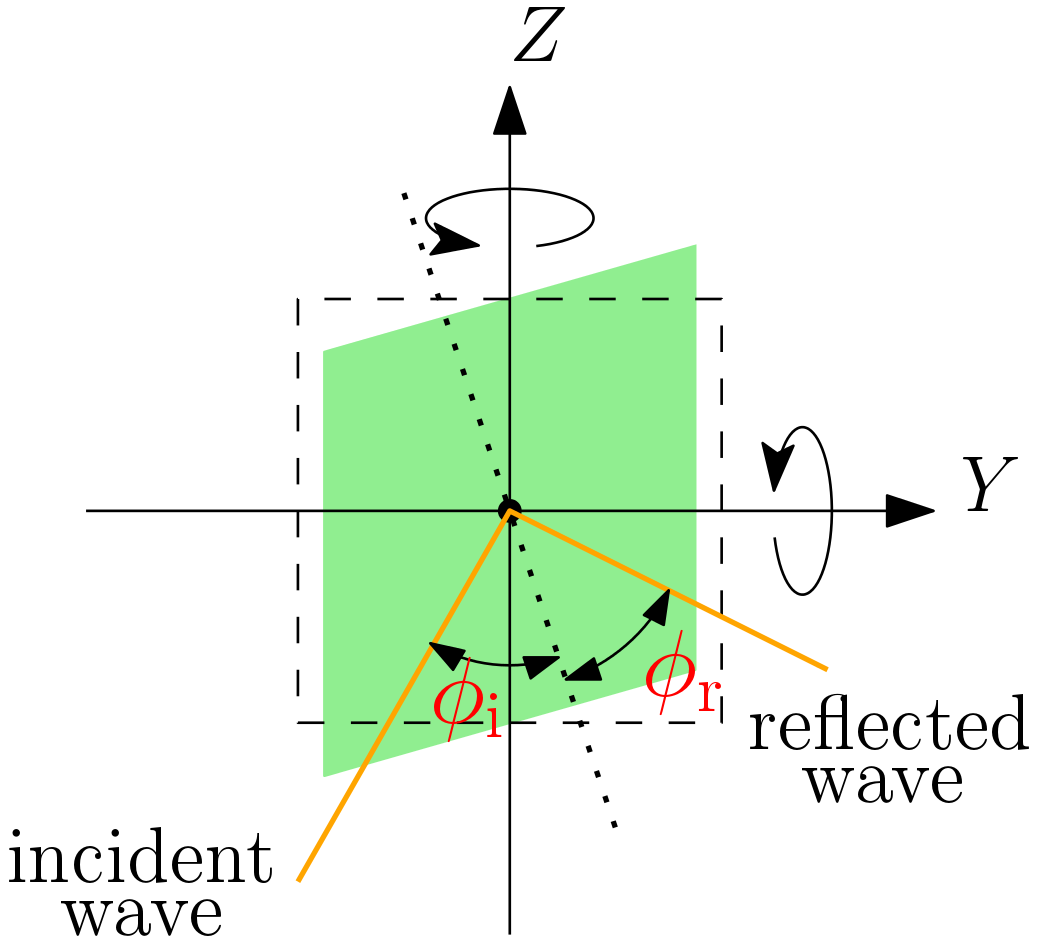}}
\caption{Waves reflection: metasurface patch versus small mirror.}
\label{fig:S2}
\end{figure}
\begin{figure*}[t]
\centering 
\captionsetup{justification=centering}
\includegraphics[width=0.75\linewidth]{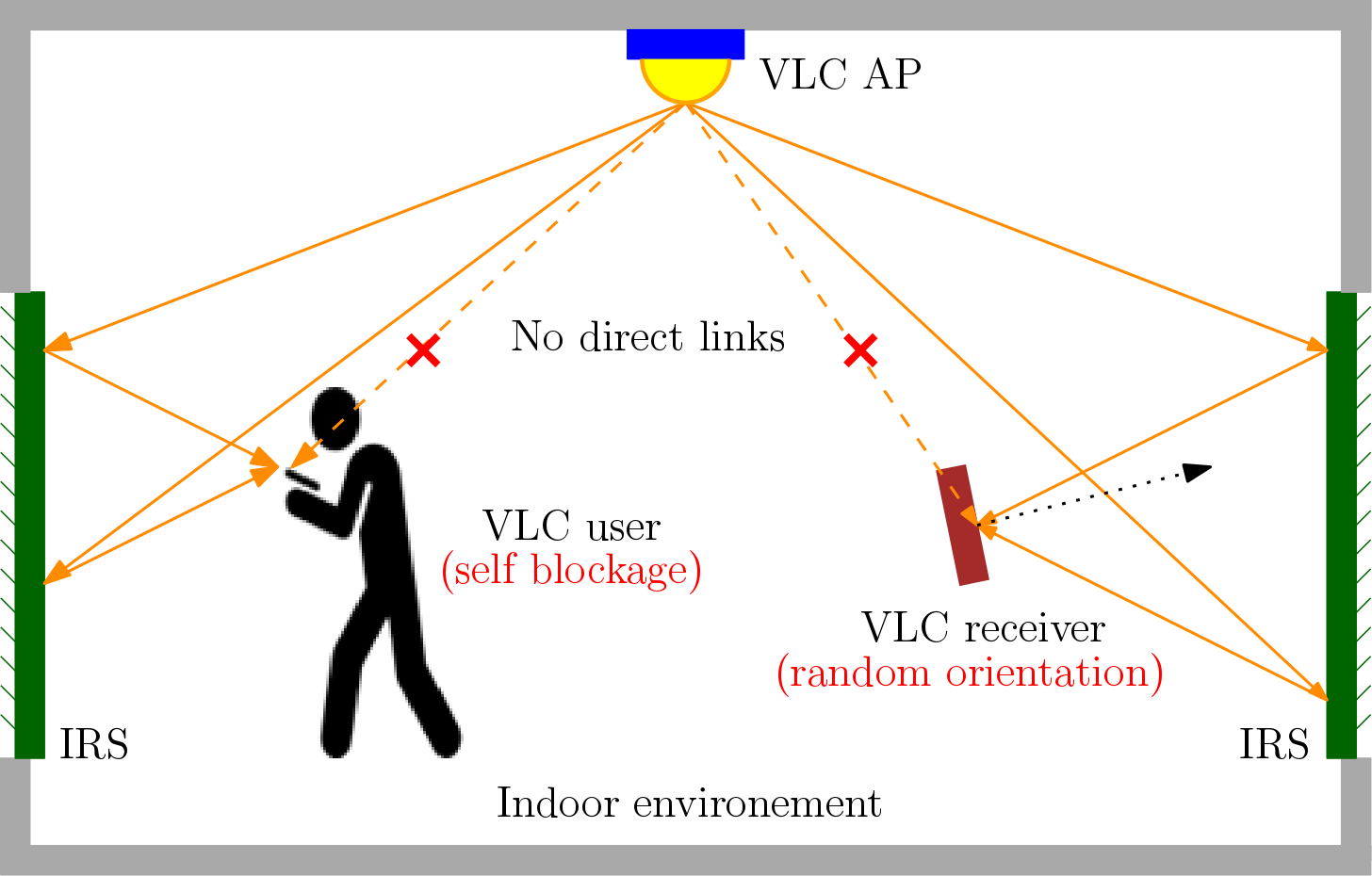}
\caption{IRS versus random orientation and link blockage.}
\label{fig:RIS_random_orientation_blockage}
\end{figure*}
\begin{enumerate}
    \item \textbf{Metasurface Array (MSA):} \\
    An MSA is a synthesized material composed of arrangements of sub-wavelength metallic or dielectric structures that are used to manipulate the light propagation in unusual ways compared to classical optical devices. This surface is capable of realizing the functionalities of many classical optical devices such as manipulating wavelength and polarization of incident waves, as well as providing new functionalities. In fact, a MSA is composed of multiple adjacent patches, where as shown in Fig.~\ref{fig:MSA_reflection}, each patch is capable of reflecting an incident optical wave toward any direction by adjusting its phase shift independently from the other patches. Such reflection is known as anomalous reflection and it is governed by the generalized Snell’s law of reflection, where by manipulating the phase gradient of the metasurface patch, the phase of the incident wave can be shifted, and therefore, the angle of reflection $\phi_{\rm r}$ can be adjusted and controlled and it is no longer equal to the angle of incidence $\phi_{\rm i}$. This kind of reflection by the metasurface patches can realize the focusing of optical power toward a VLC receiver, which can not be achieved by classical optical devices.
    \item \textbf{Mirror Array (MA):} \\
    An MA is a reflecting surface comprising multiple adjacent small mirrors that can be implemented by micro electromechanical systems. As shown in Fig.~\ref{fig:MA_reflection}, each small mirror can freely adjust its rotation, and therefore, it can manipulate the direction of an incident optical wave without changing its amplitude and polarity. Specifically, based on the principle of specular reflection, the angle of reflection $\phi_{\rm r}$ of the incident optical wave is equal to the angle of incidence. However, it can be adjusted by manipulating the rotation angles of the small mirror. Consequently, when a VLC user is at a certain position, each small mirror can adjust its rotation angles to deflect the incident optical beam toward the user. 
\end{enumerate}
\subsection{Benefits} The focusing capabilities of the MSA and MA can be exploited to control the propagation of optical beams, and therefore, in beating the effect of the random device orientation and link blockage in VLC systems. Specifically, for a certain channel use, as shown in Fig.~\ref{fig:RIS_random_orientation_blockage}, the direct LOS between certain APs and a VLC receiver may not exist due the random orientation of the VLC UE or due to blockage resulting from the presence of some random objects or from the movement of the users themselves. To overcome this, indirect LOS VLC links can be established alternatively through the use of IRS. In fact, based on the above explanation, the phase gradient of each patch in the MSA and the orientation of each small mirror in the MA can be controlled intelligently to reflect the incident optical signals toward any VLC receiver that does not have LOS links with the APs. Recently, the expression of the received power density for the case of MSA and MA were derived in \cite{abdelhady2020visible} to characterize the focusing capability of IRS in VLC systems, where it was shown that the MA outperforms the MSA.
\subsection{Example}
\begin{figure*}[t]
\centering     
\includegraphics[width=0.75\linewidth]{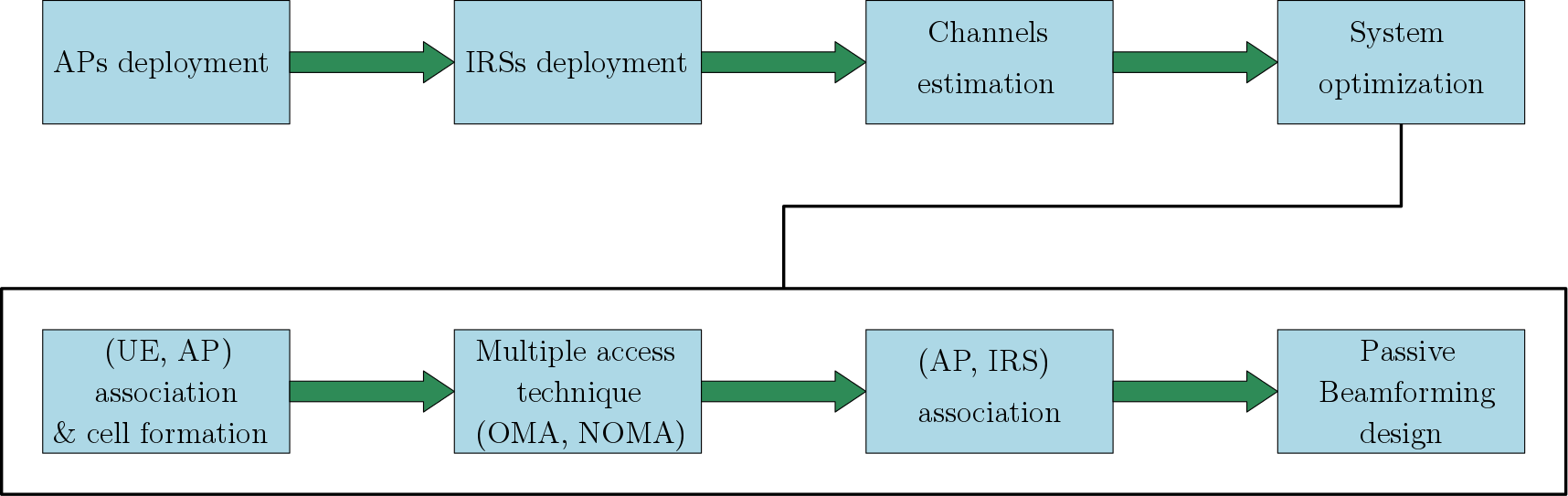}
\caption{Block diagram of an IRS-aided VLC System.}
\label{fig:block_diagram}
\end{figure*}
\begin{figure}[t]
\centering     
\includegraphics[width=1\linewidth]{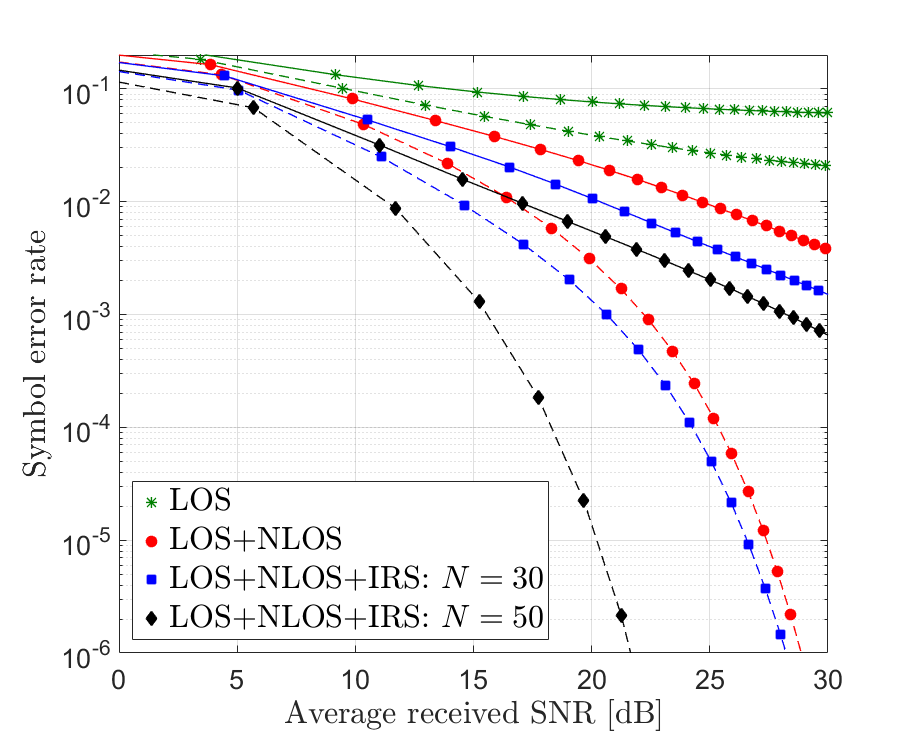}
\caption{SER of the OOK modulation versus the average received SNR. Solid lines represent the case when the density of blockers is $1$ m$^{-2}$ and dashed lines represent the case of a blockage-free environment.}
\label{fig:SER}
\vspace{-0.3cm}
\end{figure}

\indent The gain brought by integrating IRS in indoor practical VLC systems is evaluated in the following. Consider a typical indoor environment with dimensions $L\times W \times H$ = $5\times 5 \times 3$ m$^3$, where $L$, $W$ and $H$ denotes the length, the width and the height of the room. A VLC AP, consisting of only one LED, is installed at the center of the room's ceiling. Moreover, a VLC UE with a random orientation is randomly located within the indoor environment. A Monte Carlo simulation of $10^4$ independent trials is conducted, where at each trial, a random location and a random orientation for the UE are generated using the measurements-based VLC channel statistics derived in \cite{arfaoui2020measurements,mohammad2018optical}. In addition, a set of blockers is randomly generated within the indoor environment using a Poisson point process with a certain density, where the dimension of each blocker is $0.75 \times 0.2 \times 1.75$ m$^3$. Furthermore, four MAs are deployed within the indoor environment, where for each wall, only one MA is placed at its center. Each MA consists of $N \times N$ rectangular, adjacent and equally sized controllable small mirrors. The size of each small mirror is $0.1 \times 0.06$ m$^2$ and its orientation is optimized according in a way that produces the highest received optical power. \\
\indent Fig.~\ref{fig:SER} presents the symbol error rate (SER) of the on-off-keying (OOK) modulation for the three following scenarios:
\begin{enumerate}
    \item When only the LOS component of the channel gain between the AP and the UE is considered. 
    \item When both the LOS and the NLOS components (resulting from all reflections from walls) of channel gain between the AP and the UE are considered. 
    \item When the LOS and the NLOS components of the channel gain between the AP and the UE along with the cascaded channel gain resulting from the reflections by the deployed IRSs are considered.
\end{enumerate}
In accordance with the results of \cite{arfaoui2020measurements,mohammad2018optical}, the SER is significantly degraded when the NLOS components are ignored because it is highly likely that the channel gain is zero due to random orientation and link blockage. In such a case, the information will be lost, and the SER saturates at a high SNR when only the LOS component is considered. On the other hand, let us consider the SER of $3.8 \times 10^{-3}$, which is the soft forward error correction limit, as the target error performance. For the case when the indoor environment is free of blockage, and as it can be seen from Fig.~\ref{fig:SER}, the required received SNR without including the IRSs is approximately $30$ dB, whereas the required received SNR when the IRSs with $N\times N = 50 \times 50$ elements are integrated is approximately $23$ dB, i.e., a reduction of $7$ dB. Additionally, considering the case when the density of blockers is $1$ m$^{-2}$, the required received SNR without including the IRSs is approximately $23$ dB, whereas the required received SNR when the IRSs with $N\times N = 50 \times 50$ elements are integrated is approximately $16$ dB, i.e., a gain of $7$ dB can be achieved.

\section{Key Aspects of Integrating IRS in Indoor VLC Systems and Open Research Problems}
\indent Although IRS seems to have great capabilities in enhancing the performance of VLC systems, several concerns related to its integration into VLC systems have to be addressed in order to harness its full potential. Such concerns include its deployment in indoor VLC environments, CSI acquisition, configuration optimization (phase shifts and/or rotation angles) and IRS control (feedback of the preferred or the optimized configuration). These concerns will be addressed in the following subsections.
\vspace{-0.3cm}
\subsection{IRSs Deployment}
\label{deployment}
The block diagram of an IRS-aided indoor VLC system is shown in Fig.~\ref{fig:block_diagram}, and as it can be seen, the first phase is the deployment of APs that will serve for both illumination and communication purposes. One of the key features of VLC systems is the exploitation of the existing luminous infrastructure for wireless communication and data transmission. However, for newly built indoor sites, the deployment of APs should be optimized in terms of the number and locations of APs in order to guarantee simultaneously the illumination and luminosity requirements and the target performance of the integrated VLC system (number of served VLC users, coverage, quality of service (QoS), etc). Subsequently, and as shown in the block diagram in Fig.~\ref{fig:block_diagram}, the second phase is the deployment of IRSs. This phase consists of determining the optimal number and locations of IRSs with the objective of guaranteeing certain performance indicators of VLC systems, such as coverage percentage and QoS. For new indoor environments, the first and second phases can be performed jointly. In fact, optimal joint deployment of VLC APs and IRSs is better than optimizing the deployment of each one of them separately, especially in terms of cost, hardware implementation and energy efficiency. \\ 
\indent In practice, the indoor propagation environments may be complicated and several considerations should be taken into account when deploying IRSs in indoor VLC systems, including 1) the spatial user density, where high priority should be given to the zones with a large number of users; 2) the inter-cell interference (ICI). Although this problem can be solved using coordinated broadcasting between adjacent APs, some IRSs can be also deployed near the boundaries of adjacent cells to reduce the ICI effects; 3) the specifications of the indoor environment, such as the site layout along with the density and/or the locations of possible blockers and their dimensions. In fact, the deployments of IRSs is site-dependent and there is no unified deployment approach that can apply to all indoor environments; 4) some blockers may be mobile, such as human bodies and the users themselves, which render the deployment decisions even harder; and 5) in indoor environments, IRSs can be only placed on the walls and their dimensions are also limited by the dimensions of the indoor environment.  \\
\indent The deployment of IRSs in indoor environments is a crucial task \cite{wu2019towards}. Different techniques can be used to solve the problem of deployment of IRSs in indoor VLC systems, where each method has its benefits and drawbacks. First, the exhaustive search for the optimal placement of IRSs. Although this technique is optimal, it requires the CSI at all possible locations, and therefore, it is computationally costly and also require site-specific information such as building and floor layouts. Second, using some good heuristics. These approaches provides low computational complexity at the expense of the optimal of the deployment decisions. Due to this, how to achieve autonomous deployment of IRSs in indoor VLC systems by identifying their most suitable locations is a research problem of high practical interest. 
\vspace{-0.3cm}
\subsection{CSI Acquisition}
\begin{figure}[t]
\centering     
\includegraphics[width=0.75\linewidth]{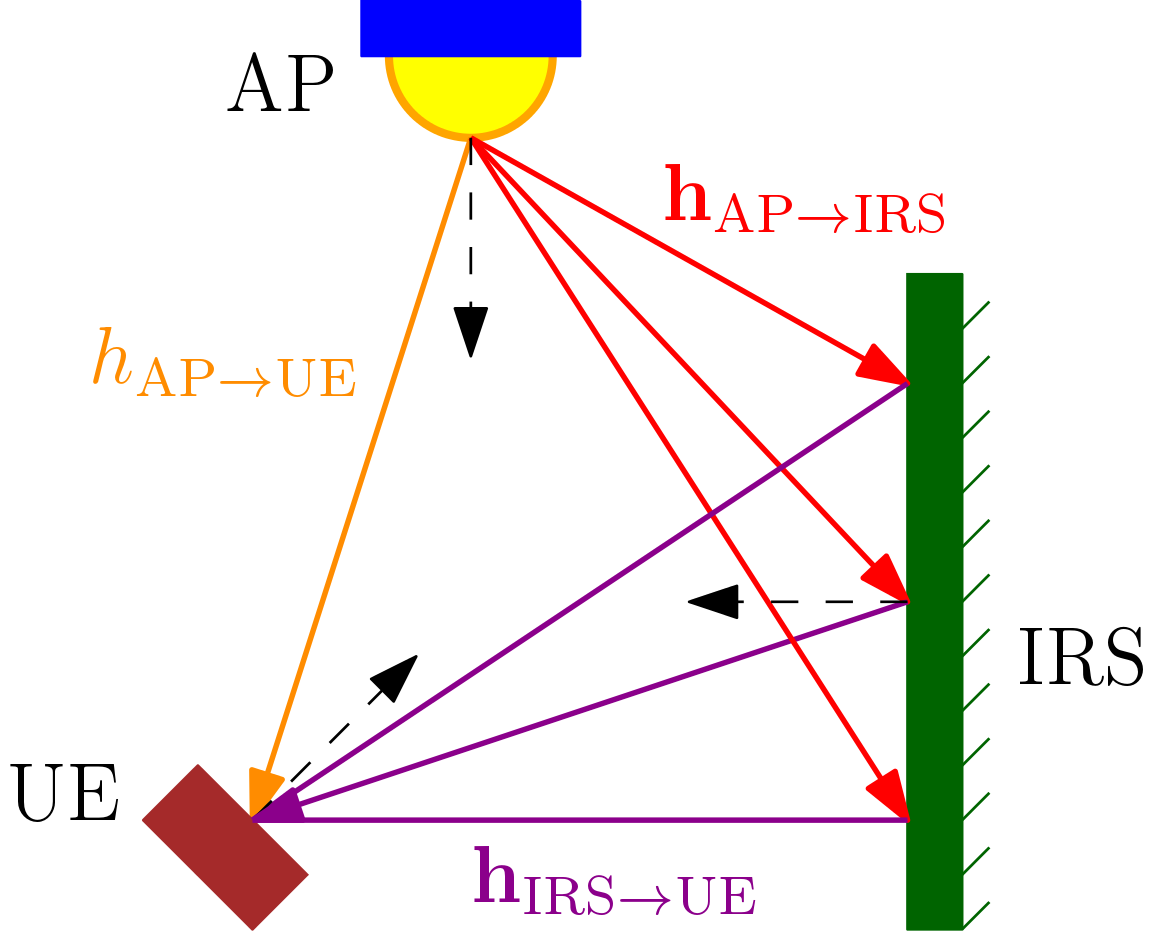}
\caption{Wireless links of an IRS-aided VLC system.}
\label{fig:channel_estimation}
\end{figure}

The second crucial point that should be addressed in IRS-aided VLC systems is the CSI acquisition, which has been identified as one of the open research problems in the literature of IRS-aided outdoor RF, mmWave and FSO cellular systems \cite{wu2019towards}. However, this problem is somehow solved in IRS-aided VLC systems as it will be explained in the following. Let us consider a case when an AP, a UE and an IRS are associated together as shown in Fig.~\ref{fig:channel_estimation}. In this context, three different wireless links can be identified. First, the channel gain $h_{\rm AP \rightarrow UE}$ between the AP and the UE, which includes the LOS and the NLOS components. Second, the channel gain vector $\mathbf{h}_{\rm AP \rightarrow IRS}$ between the AP and the IRS, which includes the channel gain between the AP and each element of the IRS. Third, the channel gain vector $\mathbf{h}_{\rm IRS \rightarrow UE}$ between the IRS and the UE, which includes the channel gain between each element of the IRS and the UE.\\
\indent The point to point VLC channel gain between any transmitter and any receiver depends mainly on the position and orientation of the transmitter and receiver. Therefore, the estimation of the channel gains defined above depends on the position and orientation of the AP, the IRS elements and the UE. In this context, once the APs are deployed, their locations and orientations are fixed and known at the VLC controller, which is the central unit responsible for controlling and operating the VLC system. A typical VLC AP is usually mounted on the ceiling of the indoor environment and its orientation is practically downward-facing. Considering the IRS elements, their exact locations are fixed and known at the VLC controller once the IRS is deployed. Moreover, for the case of a MSA, the orientation of the metasurface patches is horizontal and unchangeable, whereas for the case of a MA, the optimal orientation of each small mirror is determined by the VLC controller. Therefore, the orientations of the IRS elements are also known at the VLC controller, and based on this explanation, the channel gain vector $\mathbf{h}_{\rm AP \rightarrow IRS}$ between the AP and the IRS can be fully determined.\\
\indent Considering the position and orientation of the UE, they can be characterized through its 3D coordinates and 3D orientation angles with respect to a global coordinate system. Hence, the 3D position coordinates of the UE can be estimated at the VLC controller using indoor VLC positioning systems \cite{zhou2019joint}. Moreover, current smart devices, such as smartphones, wearables, and sensors, are equipped with an inertial measurement unit (IMU) that includes a gyroscope, an accelerometer and/or a compass. Therefore, using the IMU mounted in current smart devices, the 3D orientation angles can be estimated at the UE and then fed back to the VLC controller, using either infrared uplink transmission, which is a component of current light-fidelity (LiFi) technology \cite{haas2015lifi}, or WiFi links \cite{poulose2019indoor}. Based on this explanation, since the VLC controller knows the exact locations and orientations of the IRS elements and can estimate the position and the orientation of the UE, the channel gain $\mathbf{h}_{\rm AP \rightarrow UE}$ between the AP and the UE and the channel gain vector $\mathbf{h}_{\rm IRS \rightarrow UE}$ between the IRS elements and the UE can be estimated as well. 
\vspace{-0.3cm}
\subsection{System Optimization}
\indent Once the channel gains of the different communication links are estimated, the system performance will be optimized by the VLC controller. Assuming that a set of UEs are attempting to communicate simultaneously with the APs, the first step is to associate the UEs with the APs based on their channel gains estimated in the CSI acquisition phase. Hence, each subset of APs is responsible for transmitting the data to a subset of UEs, which leads to the formation of multiple VLC cells. Subsequently, the second step consists of selecting the multiple access technique that should be adopted within each cell, either an orthogonal multiple access technique, such as orthogonal-frequency-division-multiple-access, or a non-orthogonal-multiple-access (NOMA) technique, such as power-domain NOMA. In addition, some UEs, especially those located at the boundaries of adjacent cells, may need to be jointly served from the adjacent APs. Coordinated broadcasting can be adopted to serve these UEs and this should be decided within this step as well. \\
\indent According to the performance metric considered in the optimization process, each IRS will be associated with one or multiple APs to assist the communication to different cellular UEs. In fact, the VLC controller should decide which IRS should assist the communication from which AP and toward which subset of UEs. Obviously, this is an association problem that takes into account the CSI between the APs, the UEs and the different IRSs. Afterwards, the following step consists of optimizing the configuration of each IRS, i.e., the passive beamforming design, along with other resource allocation variables from the APs, such as power, frequency and/or time, in a way that achieves the highest performance of the VLC cellular system or that guarantees certain QoS for the cellular UEs. The optimization variables with respect to the IRS depends on its type. For the case of a MSA, the goal is to determine the optimal phase shift (or phase gradient) for each metasurface patch, whereas for the case of an MA, the goal is to determine the optimal rotation angles of each small mirror. \\
\indent While the optimization framework of VLC systems is different from that of RF systems due to its different operating constraints, the same optimization techniques can be used for obtaining the optimal IRSs configurations, such as the semidefinite relaxation and the block coordinate descent, to name a few. However, the main drawback of these techniques is their high computational complexity. In fact, for a typical IRS that is equipped with a large number of reflective elements, there is a large number of variables (phase shifts or rotation angles) that should be optimized, which results in a high computational complexity. Therefore, in order to enable real-time communication links, accurate and low complexity optimization techniques are required for this task, and it is considered indeed as a promising research direction.\\
\indent On the other hand, for the case when the IRS is serving only one UE, the IRS realizes the beam focusing function, where the optical beam reflected by the IRS is focused toward the UE. However, for the case when the IRS is serving multiple UEs simultaneously, the IRS realizes the beam splitting function. Specifically, by adjusting the phase shifts of the MSA, the incident optical beam can be split into sub-beams and each sub-beam can be focused toward a specific direction. On the other hand, the small mirrors of an MA can be grouped into different clusters and the incident optical beam can be focused toward multiple spatially dispersed UEs by selecting different clusters of small mirrors for deflection. This problem is equivalent to a clustering problem of small mirrors and it represents also a potential research direction. 
\vspace{-0.2cm}
\subsection{IRS Control}
The full potential of IRS depends heavily on the proper configuration of its elements. In this context, there is an additional critical operation that should be addressed, which is the real time control of the IRS. In fact, when the VLC controller determines the optimal IRS configuration, the question that arises here is \textit{how can this configuration be fed to the IRS in real time?} Some solutions can be adopted to deal with this problem. The first solution consists of dedicating a control channel that transmits the control signals to each IRS. Such control channel can be either optical (LiFi) or radio (WiFi). Otherwise, after deploying the IRSs in the indoor environment, and since the IRSs are basically mounted on walls, the second solution may be linking the VLC controller and each IRS controller through an optical fiber link that can communicate the control signals. 
\section{Conclusion}
In this paper, an overview of the auspicious IRS technology for boosting the performance of indoor VLC systems is provided. In typical VLC systems, the IRS configuration can be adjusted dynamically to adapt to the random behaviour of the wireless propagation environment in terms of mobility and random orientation of VLC users as well as the random blockage of optical links. This paper shows that that the integration of the IRS technology in indoor VLC systems will fundamentally change their architecture from the traditional one with only active components to a new hybrid one with both active and passive components operating in an intelligent way. This will thus open rich directions for future research.
\bibliographystyle{IEEEtran}
\bibliography{main.bib}
\vspace{-0.3cm}
\renewenvironment{IEEEbiography}[1]
  {\IEEEbiographynophoto{#1}}
  {\endIEEEbiographynophoto}
\begin{IEEEbiography}{Mohamed Amine Arfaoui} (S'16) received the M.Sc. degree in information systems engineering from Concordia University in 2017. He is currently pursuing the Ph.D. degree in information systems engineering with Concordia University. His research interests include communication theory, optical communications and physical layer security. 
\end{IEEEbiography}
\begin{IEEEbiography}{Ali Ghrayeb} (F'19) received the Ph.D. degree in electrical engineering from The University of Arizona, Tucson, AZ, USA, in 2000. He is currently a Professor with the Department of Electrical and Computer Engineering, Texas A$\&$M University at Qatar. His research interests include wireless and mobile communications, physical layer security, massive MIMO, and visible light communications. 
\end{IEEEbiography}
		
\begin{IEEEbiography}{Chadi M. Assi} (F'20) received the Ph.D. degree from the City University of New York (CUNY) in 2003. He is currently a Full Professor at Concordia University. His current research interests are in the areas of network design and optimization, network modelling, and network reliability.
\end{IEEEbiography}
\end{document}